%% file: main.tex
\documentclass[acmsmall]{acmart}

\usepackage{amsmath,amsfonts}
\usepackage{lipsum}
\usepackage{algorithmic}
\usepackage{graphicx}

\usepackage{enumitem}
\usepackage{textcomp}
\usepackage{longtable}
\usepackage{float}
\usepackage{graphicx}
\usepackage{multirow}
\usepackage{balance}
\usepackage{hyperref}
\usepackage{tabularx} 
\usepackage{booktabs} 
\usepackage{colortbl}
\usepackage{hhline}
\usepackage{array}
\usepackage{subfigure}
\usepackage{xcolor}
\usepackage{enumitem}

\pdfoutput=1

\usepackage{graphicx}
\usepackage{caption}
\usepackage[inkscapearea=page]{svg}

\definecolor{coolblack}{rgb}{0.0, 0.18, 0.59}

\definecolor{background_gray}{gray}{0.85}
\definecolor{background_green}{rgb}{7,163,90}

\newenvironment{boxedtext}
    {    
    \begin{center}
    
    \begin{tabular}{|p{0.96\linewidth}|}
    \hline
    }
    { 
    \\ \hline
    \end{tabular} 
    
    \end{center}
       }

\definecolor{MidnightBlue}{rgb}{0.1, 0.1, 0.44}

\newcolumntype{L}[1]{>{\raggedright\let\newline\\\arraybackslash\hspace{0pt}}m{#1}}
\newcolumntype{C}[1]{>{\centering\let\newline\\\arraybackslash\hspace{0pt}}m{#1}}
\newcolumntype{R}[1]{>{\raggedleft\let\newline\\\arraybackslash\hspace{0pt}}m{#1}}

\def\BibTeX{{\rm B\kern-.05em{\sc i\kern-.025em b}\kern-.08em
    T\kern-.1667em\lower.7ex\hbox{E}\kern-.125emX}}

\AtBeginDocument{%
  \providecommand\BibTeX{{%
    \normalfont B\kern-0.5em{\scshape i\kern-0.25em b}\kern-0.8em\TeX}}}

\setcopyright{cc}
\setcctype{by}
\acmDOI{10.1145/3715744}
\acmYear{2025}
\acmJournal{PACMSE}
\acmVolume{2}
\acmNumber{FSE}
\acmArticle{FSE029}
\acmMonth{7}
\received{2024-09-13}
\received[accepted]{2025-01-14}

\begin{document}

\title{The Landscape of Toxicity: An Empirical Investigation of Toxicity on GitHub}

\author{Jaydeb Sarker}
\orcid{0000-0001-6440-7596}
\affiliation{%
  \institution{University of Nebraska at Omaha}
  \city{Omaha}
  \country{USA}
}
\email{jsarker@unomaha.edu}

\author{Asif Kamal Turzo}
\orcid{0000-0002-0869-4962}
\affiliation{%
  \institution{Wayne State University}
  \city{Detroit}
  \country{USA}
}
\email{asifkamal@wayne.edu}

\author{Amiangshu Bosu}
\orcid{0000-0002-3178-6232}
\affiliation{%
  \institution{Wayne State University}
  \city{Detroit}
  \country{USA}
}
\email{amiangshu.bosu@wayne.edu}

\input{Sections/abstract}



\maketitle

\input{Sections/introduction}
\input{Sections/related-works}

\input{Sections/methods}

\input{Sections/results}
\input{Sections/discussion}
\input{Sections/threats}

\input{Sections/conclusion}

\input{Sections/acknowledgement}

\input{Sections/data_avail}

\bibliographystyle{ACM-Reference-Format}  

\bibliography{bibliography}

\end{document}

%% file: Sections/abstract.tex
\begin{abstract}
Toxicity on GitHub can severely impact Open Source Software (OSS) development communities. To mitigate such behavior, a better understanding of its nature and how various measurable characteristics of project contexts and participants are associated with its prevalence is necessary. To achieve this goal, we conducted a large-scale mixed-method empirical study of 2,828 GitHub-based OSS projects randomly selected based on a stratified sampling strategy. Using ToxiCR, an SE domain-specific toxicity detector, we automatically classified each comment as toxic or non-toxic. Additionally, we manually analyzed a random sample of 600 comments to validate ToxiCR's performance and gain insights into the nature of toxicity within our dataset. The results of our study suggest that profanity is the most frequent toxicity on GitHub, followed by trolling and insults. While a project's popularity is positively associated with the prevalence of toxicity, its issue resolution rate has the opposite association. Corporate-sponsored projects are less toxic, but gaming projects are seven times more toxic than non-gaming ones. OSS contributors who have authored toxic comments in the past are significantly more likely to repeat such behavior. Moreover, such individuals are more likely to become targets of toxic texts.
\end{abstract}

%% file: Sections/introduction.tex
{\small \textcolor{red}{ Warning: This paper contains examples of language that some people may find offensive or upsetting.} }

\section{Introduction}
\label{sec:intro}
In 2023, GitHub, the most popular Open Source Software (OSS) project hosting platform, hosted more than 284 million public repositories~\cite{octoverse-2023}.  As OSS communities continue to grow, so do the interactions among contributors during various software development activities, such as issue discussions and pull request reviews. 
While these interactions are crucial to facilitating collaborations among the contributors, a few may take negative turns and cause harm~\cite{GitHub_GitHub_Open_Source}. 
Recent studies have investigated the umbrella of antisocial interactions among OSS developers using various lenses, which include  `toxicity,'~\cite{miller2022did, sarker2022automated,sarker2020benchmark,raman2020stress,sarker-esem-2023}, `incivility'~\cite{ferreira2021shut}, `destructive criticism'~\cite{gunawardena2022destructive}, and `sexism and misogyny'~\cite{sultana2021rubric}. Although these lenses differ, they all share a common attribute: the potential to cause severe repercussions among the participants. Consequences of antisocial interactions include stress and burnout~\cite{raman2020stress}, negative feelings~\cite{egelman2020predicting,ferreira2021shut,gunawardena2022destructive}, pushbacks~\cite{murphy2022pushback}, turnovers of long-term contributors~\cite{toxic-blog-linux2, toxic-open-source-maintainer, perl-toxic-2,leaving-for-toxicity}, adding barriers to newcomers' onboarding~\cite{raman2020stress}, and hurting diversity, equity, and inclusion (DEI) by disproportionately affecting women and other underrepresented minorities~\cite{gunawardena2022destructive,albusays2021diversity,nafus2012patches,murphy2022pushback}. Moreover, the prevalence of antisocial behaviors present substantial challenges to the growth and sustainability of an OSS project.

Therefore, recent Software Engineering (SE) research has focused on characterizing {toxicity and other} antisocial behaviors and their consequences through surveys, interviews, and qualitative analyses~\cite{ferreira2021shut,miller2022did,egelman2020predicting,gunawardena2022destructive,ferreira2022heated}. 
In a sample of 100 GitHub issue comments, ~\citet{miller2022did} found entitlement, arrogance, insult, and trolling as the most common forms of toxicity. Destructive criticism is another anti-social behavior found in code reviews~\cite{gunawardena2022destructive}. Although destructive criticisms are rare, they may have severe repercussions, which include conflicts, demotivation, and even hindering the participation of minorities~\cite{gunawardena2022destructive, egelman2020predicting}. 
~\citet{ferreira2021shut}'s study of rejected patches in Linux kernel mailing lists reported frustration, name-calling, and impatience as the most prevalent forms of incivility. Another recent workplace investigation reported inappropriate communication style as the primary cause of incivility~\cite{rahman2024words}.  
However, many of these studies suffer from limitations such as small sample sizes or narrow focuses on specific projects~\cite{ferreira2022heated,miller2022did},  organizations~\cite{qiu2022detecting,egelman2020predicting}, or a small developer group~\cite{rahman2024words}, which raises questions regarding the external validity of these findings at different contexts.  We also lack a quantitative empirical investigation of how various measurable characteristics of project contexts and participants are associated with the prevalence of antisocial behavior, such as toxicity.  
Such an investigation is necessary to formulate context-aware toxicity mitigation strategies for the broader OSS ecosystem.

In response to this need, we have conducted a large-scale empirical investigation of toxicity during  Pull Requests (PRs). 
We selected PR since it is a crucial mechanism to attract contributions from non-members and facilitate newcomers' onboarding among OSS projects~\cite{gousios2014exploratory}. PRs allow contributors to propose changes, which other community members then review PRs. Due to the interpersonal nature of PR interactions and the potential for dissatisfaction due to unfavorable decisions, PR interactions may raise conflicts and anti-social behaviors. We select the `toxicity' lens since it has been most widely investigated~\cite{raman2020stress,miller2022did,sarker-esem-2023,sarker2020benchmark} and has a reliable automated identification tool~\cite{sarker2022automated}, which is a prerequisite for a large-scale empirical investigation. 
 Following the toxicity investigation framework proposed by \citet{miller2022did}, we investigate four research questions to characterize: i) the nature of toxicity, ii) projects that had higher toxic communication than others, iii) contextual factors that are more likely to be associated with toxicity, and iv) the participants of toxic interactions, respectively. We briefly motivate each of the research questions as follows.

\vspace{2pt}
\noindent \textbf{RQ1:  [Nature]} \textit{What are the common forms of toxicity observed during GitHub Pull Requests?}

\textit{Motivation:} 
Understanding the nature of toxicity may help project maintainers improve guidelines and interventions to foster respectful and constructive interactions. Prior studies have proposed various categorizations of SE domain-specific toxicity~\cite{miller2022did, sarker2022automated} and incivility~\cite{ferreira2021shut}. However, due to sampling criteria used in those studies (e.g., only locked issues, small samples, or a specific project), two key insights remain under-explored: i) whether these studies missed additional forms of toxicity, and ii) how frequently various toxicity categories occur on GitHub. RQ1 aims to fill in these knowledge gaps.

\vspace{2pt}
\noindent  \textbf{RQ2: [Projects]} \textit{What are the characteristics of the project that are more likely to encounter toxicity?}

\textit{Motivation:}  Does toxicity vary across project sponsorship, age,  popularity, quality, domain, or community size?  The identification of these factors will help project management undertake context-specific mitigation strategies. Determining which projects are more likely to suffer from toxicity may allow a project's management to allocate resources and define mitigation strategies.  Moreover, this insight will help a prospective joiner select projects.

\vspace{2pt}
\noindent \textbf{RQ3: [PR Context]} \textit{Which pull requests are more likely to be associated with toxicity on GitHub?}

\textit{Motivation:}
Does toxicity occur during poor-quality changes, unfavorable decisions, large changes, or delayed decisions?
Understanding contextual factors is crucial to educating developers to avoid creating specific scenarios.

\vspace{2pt}
\noindent \textbf{RQ4: [Participants] } \textit{What are the characteristics of participants associated with toxic comments?}

\textit{Motivation:}
Prior studies~\cite{miller2022did,ferreira2021shut} suggested that some participants are likelier to author toxic comments due to their communication style or cultural background. On the other hand, another study suggests that participants representing underrepresented groups or newcomers are more likely to be targets~\cite{murphy2022pushback}. RQ4 aims to identify personal characteristics associated with being authors or victims of toxicity. This insight will help project management prepare community guidelines to combat toxicity and protect vulnerable participants.

\vspace{2pt}
\noindent \textbf{Research method:} We conducted a large-scale mixed-method empirical study of 2,828 GitHub-based OSS projects randomly selected based on a stratified sampling strategy. Our sample includes 16 million PRs and 101.5 million PR comments.  Using ToxiCR~\cite{sarker2022automated}, a state-of-the-art SE domain-specific toxicity detector, we automatically classified each comment as toxic or non-toxic. Additionally, we manually analyzed a random sample of 600 comments to validate ToxiCR's performance and gain insights into the nature of toxicity within our dataset. With ToxiCR demonstrating a reliable performance, we trained multivariate regression models to explore the associations between toxicity and various attributes of projects, PR contexts, and participants.

\vspace{2pt}
\noindent \textbf{Key findings:}
We found 11 forms of toxic comments among GitHub PR reviews, with object-directed toxicity being a new form unreported in prior studies. 
The results of our study suggest that profanity is the most frequent toxicity on GitHub, followed by trolling and insults. 
While a project's popularity is positively associated with the prevalence of toxicity, its issue resolution rate has the opposite association.
Corporate-sponsored projects are less toxic, but gaming projects are seven times more toxic than non-gaming ones. OSS contributors who have authored toxic comments in the past are significantly more likely to repeat such behavior. Moreover, such individuals are more likely to become targets of toxic texts.

\vspace{4pt}
\noindent \textbf{Contributions} The primary contributions of this study include:
\begin{itemize}
\item  An empirical investigation of various categories of toxic communication among GitHub {PRs}.
\item  A large-scale empirical investigation of factors associated with toxicity on GitHub.
\item Actionable recommendations to mitigate toxicity among OSS projects. 

\item  We publicly make our dataset and scripts available at: \hyperlink{https://doi.org/10.5281/zenodo.14802294}{10.5281/zenodo.14802294}~\cite{dataset}.
 
\end{itemize}

\vspace{2pt}
\noindent \textbf{Organization:} The remainder of the paper is organized as follows.
Section~\ref{sec:background} briefly overview closely related works.
Section~\ref{sec:research-method} details our research methodology.  
 Section~\ref{sec:results} presents the results of our empirical investigation.  
 Section~\ref{sec:discussion} and Section~\ref{sec:threats} discuss implications and threats to the validity of our findings, respectively. 
 Finally, Section~\ref{sec:conclusion} concludes the paper.

%% file: Sections/related-works.tex
\section{Related Works}
\label{sec:background}
\textbf{Antisocial Behaviors in OSS:} While internet-based communication mediums help people across multiple geographical regions to collaborate with ease, these interactions can turn negative due to various anti-social behaviors such as toxicity~\cite{bhat2021say}, harassment~\cite{lindsay2016experiences}, cyberbullying~\cite{kowalski2014bullying}, trolling, and hate speech~\cite{del2017hate, gagliardone2015countering}.
Although these behaviors are less frequent among professional work-focused communities such as OSS projects than social mediums~\cite{miller2022did}, they may have severe repercussions on the productivity and even the sustainability of an OSS project~\cite{raman2020stress,toxic-open-source-maintainer,perl-toxicity}.
Among the various lenses studied by SE researchers, 
`toxicity,' which is `behaviors that are likely to make someone leave,' has been investigated most frequently~\cite{raman2020stress, sarker2020benchmark, sarker2022automated, qiu2022detecting}. Toxicity among OSS projects on GitHub differs from other social communication platforms 
such as Reddit, Wikipedia, Twitter, and Stack Overflow~\cite{miller2022did}. 
`Incivility,' defined as a broader superset including toxicity, is a text with an unnecessary disrespectful tone~\cite{ferreira2021shut}.
`Destructive criticism'  is another lens characterized by negative feedback during code reviews~\cite{gunawardena2022destructive}. Interactions during code reviews may also cause interpersonal conflicts among the parties, which can be termed as `pushback'~\cite{murphy2022pushback, egelman2020predicting}. 

\vspace{3pt}
\noindent \textbf{Automated identification of toxicity and other anti-social behaviors:}
To identify and mitigate online toxic interactions, researchers from the Natural Language Processing (NLP) domain have published datasets and classifiers, where several come from Kaggle challenges \cite{zaheri2020toxic,bhat2021say,kumar2021designing,zhao2021comparative}. For the SE domain,  \citet{raman2020stress} proposed the first customized toxicity detector. However, this classifier performed poorly during subsequent benchmarks~\cite{sarker2020benchmark, miller2022did, qiu2022detecting}. ~\citet{qiu2022detecting}'s classifier aims to detect interpersonal conflicts by combining pushback~\cite{egelman2020predicting} and toxicity~\cite{raman2020stress} datasets. 
\citet{cheriyan2021towards} proposed an offensive language detector that considers a subset of toxicity, such as swearing or cursing. ~\citet{sarker2022automated} developed a rubric for toxicity in the SE domain and developed a toxicity classifier (ToxiCR) with their manually labeled 19,651 code review texts, which achieved 88.9\% F1-score for the toxic class. Two recent studies have proposed classifiers to identify uncivil comments, where \citet{ferreira2024incivility} used their dataset of locked issue comments and ~\citet{rahman2024words} augmented ToxiCR dataset with ChatGPT generated instances to improve identification of mockery and flirtation.

\vspace{3pt}
\noindent \textbf{Contexts and consequences of anti-social behaviors among OSS:}
Prior SE studies investigated contexts and consequences of anti-social behaviors using surveys and qualitative analyses. 
These studies suggest toxic interactions among OSS developers as a `poison'  that not only degrades their mental health but~\cite{carillo2016towards} also can cause stress and {burnout}~\cite{raman2020stress}.  
The threat of an OSS community disintegrating rises with the levels of toxicity due to developers' turnover~\cite{carillo2016towards}. 
\citet{miller2022did}'s investigation found toxicity originated from both newcomers and experienced contributors due to various causes, which include technological disagreements, frustrations with a system, and past interactions with the target.
Project sponsorship and domain may influence toxicity as corporate projects are less toxic than non-corporate projects. On the other hand, gaming projects are more toxic than non-gaming ones~\cite{raman2020stress}. A project's toxicity may also decrease with age~\cite{raman2020stress}.
While uncivil discussions may arise in various locked issue contexts, they are more common among versioning and licensing discussions~\cite {ferreira2022heated}. On the Linux kernel mailing list, inappropriate feedback from maintainers and violation of community conventions are the top causes of incivility~\cite{ferreira2021shut}. On the other hand, among industrial developers, excessive workloads and poor-quality code are top factors~\cite{rahman2024words}.
Two lenses of antisocial behaviors, destructive criticism, and pushback, are specific to code reviews. They occur due to unnecessary harsh critiques of code and interpersonal conflicts caused by disagreements over development directions~\cite{egelman2020predicting,murphy2022pushback,gunawardena2022destructive}. 
Both pushback and destructive criticisms not only decrease productivity and degrade interpersonal relationships~\cite{murphy2022pushback,egelman2020predicting}, they disproportionately harm underrepresented minorities and cause barriers to promoting DEI~\cite{gunawardena2022destructive}.
Besides these academic works, several gray literature have also documented {burnout} and turnover of long-term OSS contributors due to toxicity~\cite{toxic-blog-linux1,toxic-blog-linux2,toxic-blog-linux3,toxic-open-source-maintainer,perl-toxicity,leaving-for-toxicity,perl-toxic-2}.

\vspace{3pt}
\noindent \textbf{Novelty:} This study differs from prior empirical investigations of antisocial behaviors in three ways. First, prior studies focused on communication from specific contexts
such as locked issues~\cite{miller2022did,raman2020stress} or rejected patches~\cite{ferreira2021shut}. Therefore, characteristics of toxicity outside these known negative contexts are missing. Second, these investigations are qualitative. While these investigations are crucial to forming hypotheses, whether these hypotheses apply to a broader spectrum of OSS projects remains unanswered. Finally, these studies explored a limited set of factors, whether other plausible factors, such as community size, project popularity, code complexity, and unresolved defects associated with toxicity, remain unanswered. For example, we investigated the association between toxicity and 32 different factors, where 21 are unique to our study.

%% file: Sections/methods.tex
\section{Research Method}
\label{sec:research-method}
\begin{figure}[t!]
	\centering  
\includegraphics[width=1\linewidth, trim=0 290 0 0, clip]{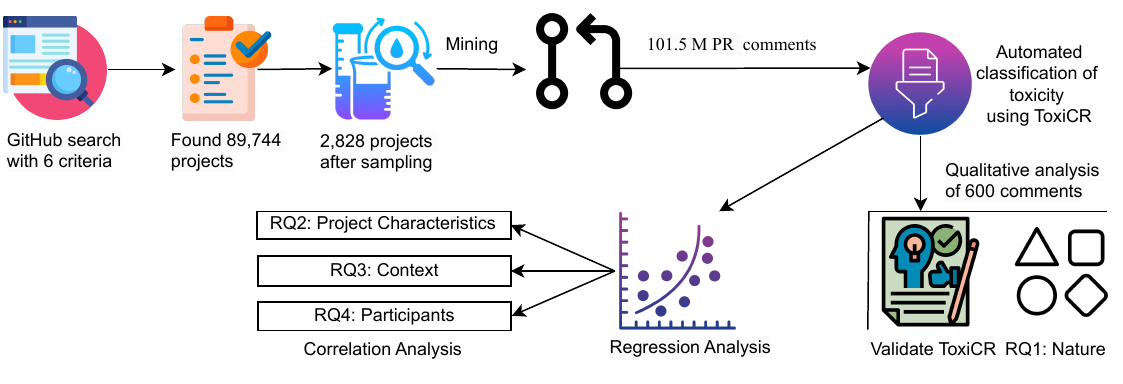}
	\caption{An overview of our research method}
	\label{fig:method_overview}	
\end{figure}

Figure~\ref{fig:method_overview} provides an overview of our research methodology, detailed in the following subsections.

\subsection{Project Selection}
We leverage the GitHub search tool developed by ~\citet{Dabic:msr2021data}, which enables filtering based on various criteria such as the number of contributors, programming language, forks, commits, and stars to select candidate projects. Following recommendations by~\citet{kalliamvakou2016depth}, we searched for projects satisfying the following six criteria: i) uses one of the top ten programming languages on GitHub:  Java, C, C++, Python, JavaScript, C{\#}, Go, PHP, Typescript, and Ruby; ii) has at least 20 contributors, iii) is publicly available with an OSS license; iv) at least two years old, v) has at least 20 PRs; vi) and has at least 10 stars. 
The first five criteria ensure the selection of OSS projects with adequate analyzable interpersonal communication among the contributors. 
The last criterion reduces the search space; without this filter, the number of projects grows exponentially but adds only trivial OSS projects.

Our search conducted in September 2022 found 89,744 projects. We exported the results as a CSV file and categorized the projects into the following three groups based on project activity, which we measured using monthly Pull Request Frequency (PRF) on GitHub. 
\begin{itemize}
    \item \emph{Low PRF (PRF-L)}: A project belongs to this group if it has less than 8 PRs per month (i.e., less than 2 PRs/week). Total 66,791(~74.5\%) projects belong to this group. 
    \item  \emph{Medium PRF (PRF-M)}: A project belongs to this group if it has between 8 -31 PRs per month (i.e., 2-8 PRs/week). Total 14,694  (16.3\%) projects belong to this group. 
    \item \emph{High PRF (PRF-H)}:  A project belongs to this group if it has more than 32 PRs per month (i.e., $>$8 PRs/week). Total 8,259  (9.2\%) projects belong to this group.
\end{itemize}

These two chosen grouping thresholds, 8 and 32, represent approximately 75 and 90 percentiles based on PRF. We chose PRF instead of commit frequency because we noticed that many popular projects use GitHub as a mirror (e.g., the Linux kernel) while development activities predominantly use other platforms. 
Next, we randomly selected 800 projects from each of the three groups.
 This stratified sampling is necessary since three-fourths of the projects found based on our search criteria belong to the PRF-L  group. Hence, a random selection will have been dominated by PRF-L projects and will fail to adequately capture the characteristics of the remaining two groups. Our sample size of 800 is adequate to obtain results within a 3\% margin of error with a 95\% confidence interval~\cite{cochran1977sampling}.

Since prior studies have suggested higher occurrences of toxicity among gaming projects~\cite{miller2022did}, we also added projects with the topic `game' and at least 10 stars. The search result found 6.1k projects. However,  most projects did not satisfy criteria such as a minimum number of participants or PRs. Hence, the gaming group includes 439 projects, adequate for  95\% confidence interval (CI) and a 5\% Margin of Error~\cite{cochran1977sampling}. We include the gaming projects in our RQ2 analysis only to investigate project characteristics associated with toxicity. However, some projects were no longer available for mining (e.g., deleted or moved). Therefore, our final dataset contains 2,828 projects.

\subsection{Dataset Preparation}

 We wrote a Python Script using the PyGithub library~\cite{pygit} to mine all the PR details, metadata, PR labels, user information, inline review comments, and PR discussions from the selected projects and store them in a MySQL database. Our data mining started in October 2022 and completed in January 2023. Our dataset requires approximately 172 GB of storage.
Our dataset includes approximately 16.1 million (M) PRs, 101.5 M comments, and 1.3 M unique users.  We also mined all publicly available user information, including profile photos, emails, full names, and user types. We exclude all the interactions from Bot accounts (i.e., userType=`Bot'). However, we also noticed many bot accounts using incorrect flags (i.e., userType =`User'). Therefore, we filtered accounts with bot-specific keywords (bot, robot, auto, Jenkins, static, etc.) in user names and full names. We manually inspected the filtered accounts to make a final determination.

\subsection{Toxicity Classification Scheme}
\label{sec:classification-scheme}
Multiple recent studies have investigated toxicity and other antisocial behaviors within OSS communities~\cite{sarker2022automated,miller2022did,rahman2024words,raman2020stress,ferreira2021shut,sultana2022identifying,gunawardena2022destructive,egelman2020predicting}. These studies enumerate several prevalent forms of such behaviors and conclude that toxicity in the OSS context are different from a non-SE domain such as social media~\cite{miller2022did}. To answer RQ1, we focused on aggregating various categories of anti-social behaviors to prepare our manual labeling scheme. We started with consolidating categories derived from studies on `toxicity'~\cite{raman2020stress,miller2022did,sarker2022automated,qiu2022detecting}. During this process, we identified overlapping concepts based on definitions included in those papers and merged those into a single category. We noticed a conflict as `self-deprecation' was marked as non-toxic by~\citet{sarker2022automated}, while~\citet{miller2022did} marks `self-directed pejorative', a similar concept as toxic. We follow ~\citet{sarker2022automated}'s definition, marking such texts as toxic only if they involve profanity since we use their tool.
Additionally,~\citet{ferreira2021shut}'s incivility lens, which encompasses a broader spectrum of anti-social behaviors, including toxicity, also includes frustration, impatience, irony, mocking, name-calling, threat, and vulgarity. Based on their definitions, `name-calling, threat, mocking, and vulgarity' overlap with existing categories identified by ~\citet{miller2022did} and ~\citet{sarker2022automated}. As such, these were considered toxic. Although `irony,' `frustration,' and `impatience' are not a part of these toxicity schemes, they may fit existing categories, such as trolling, arrogance, and insult, depending on context. It's worth noting that existing SE studies have used different terminologies to study these subjective social constructs, and an agreed-upon standard scheme or definition is currently missing and perhaps challenging to establish. 
At the end of this step, we prepared our manual labeling scheme of 10 categories (Table~\ref{tab:toxicity_types}) with a broader definition for each group.

\subsection{Automated identification of toxic comments}
\label{sec:validation-toxicr}
  We select ToxiCR~\cite{sarker2022automated} for automated classification since it is i) trained on large-scale training data, ii) developed as a reusable standalone tool, iii) is publicly available on GitHub, iv) provides a well-defined interface to conduct a large-scale classification required by this study,  v) trained on Code review data which is similar to pull requests that this study aims to analyze, and vi) reports the best performance according to its evaluation with $95.8\%$ accuracy, $90.7\%$ precision, $87.4\%$ recall, and an $88.9\%$ F1-score. ToxiCR provides the toxicity probability of a text from 0 to 1, and its authors recommend using a threshold $>=0.5$ to consider a text as toxic. Using ToxiCR's best-performing configuration~\cite{sarker2022automated}, we classified all the PR comments, totaling 101.5 million. ToxiCR found approximately 756K toxic comments (0.74\%) from our dataset.

\emph{Evaluation of ToxiCR:} Prior research on SE domain-specific NLP tools~\cite{novielli2018benchmark,novielli2021assessment} recommends independent assessments before application on new settings. Therefore, we conducted an empirical evaluation to assess ToxiCR's reliability on our dataset. To achieve this goal, we randomly selected 600 PR comments marked as toxic by ToxiCR. This sample adequately provides results within a 2.6\%  margin of error and 95\%  confidence interval~\cite{cochran1977sampling}.  We also use this sample to investigate the frequencies of various forms of toxicity on GitHub (Section~\ref{sec:manual}).

\emph{Precision:}
Two raters independently labeled those 600 samples as toxic or non-toxic and resolved the conflicts after a discussion. To mitigate the bias of the labeling process, two labelers follow the toxicity rubric from~\citet{sarker2022automated}. The agreement between the two labelers is 95.8\%, and Cohen’s kappa~\cite{cohen1960coefficient} value is \emph{$\kappa$ = 0.80}, which is `substantial'.  After conflict resolution, 532 comments were labeled toxic, suggesting $88.8\%$  precision.  
This result is within the margin of error (i.e., 2.6\%) of ToxiCR's claimed precision (90.7\%)~\cite{sarker2022automated}.

\emph{Recall:}
Evaluation of recall is also essential to ensure that ToxiCR does not miss many positive instances. On this goal, we focused on finding existing labeled toxicity datasets curated from GitHub issue requests.   
We did not use~\citet{raman2020stress}'s dataset since it has only 106 toxic instances.
We chose ~\citet{ferreira2022heated}'s dataset of locked GitHub issues, which includes 896 uncivil sentences out of 1,364.  According to ~\citet{ferreira2021shut}, incivility is a super-set of toxicity. While all toxic comments are uncivil, some uncivil comments (e.g., irony and impatience) may not fit the toxicity lens. To encounter this challenge, two authors relabeled the 896 uncivil comments based on ~\citet{sarker2022automated}'s toxicity labeling rubric. The raters achieved an inter-rater agreement of $\kappa=0.76$ and resolved the conflicts through a mutual discussion. On this dataset, ToxiCR achieved 87\% recall, which is also within the sampling margin of error of ToxiCR's reported recall (i.e., 87.4\%).

\subsection{Manual Categorization of Toxic Comments}
\label{sec:manual}
Using the 10 class classification scheme described in Section~\ref{sec:classification-scheme}, two of the authors independently placed the 532 toxic comments identified during ToxiCR's evaluation (Section~\ref{sec:validation-toxicr}) into one or more groups. We also included the `Others' category to label toxic comments that do not fit the existing ten categories. We measured the inter-rater reliability of this multiclass labeling using Krippendorff's alpha, which was 0.35, indicating a `Fair' agreement. We noticed higher ratios of disagreements since, theoretically, the number of possible labeling for a single instance is $2^{11}$. Conflicting labels were resolved through mutual discussions.
After conflict resolution, the raters reviewed the 34 instances from the `Others' group to identify missing categories. The new category identified is `Object-Directed Toxicity', which includes anger, frustration, or profanity directed toward software, products, or artifacts. 
For example, \textcolor{brown}{``also the mask sprite is beyond horrid, I might have something that could do better..''} represents this form.
With this category, they went through the labeled instances again to identify other cases that may also fall under this category since a text can fall under multiple categories. We found a total of 49 instances belonging to this new category.

\begin{table*}
    \caption{The list of attributes selected to investigate their association with project characteristics (RQ2), Pull request context (RQ3), and participants' characteristics (RQ4). We selected this set of attributes since prior studies on code reviews and anti-social behaviors suggest the likelihood of association with toxicity or conflict-instigating contexts. * -indicates attributes that were investigated in prior studies.}
    \label{tab:rq-attributes}
    \centering
    \vspace{-12pt}
    \input{Sections/Tables/combined_attributes}
\end{table*}

\subsection{Attribute Selection}
Table~\ref{tab:rq-attributes} lists attributes selected to answer RQ2, RQ3, and RQ4 introduced in Section \ref{sec:intro}. In addition to each attribute's definition, Table~\ref{tab:rq-attributes} hypothesizes why an attribute may be associated with toxicity.

\emph{RQ2: Project }
 We select eleven project characteristics attributes based on prior studies on toxicity and incivility~\cite{raman2020stress,miller2022did,ferreira2022heated,ferreira2021shut}. These 11 attributes characterize a project's activity, popularity, domain, governance, and age. 

\emph{RQ3: PR Context }
We select nine contextual attributes based on prior studies~\cite{miller2022did,thongtanunam2017review,raman2020stress,sultana2022code,egelman2020predicting,rahman2024words}. 
These attributes characterize the type of change, outcome, complexity, required review /resolution efforts, completion time, and number of identified issues in a PR.

 \emph{RQ4: Participant}
 We select six participant attributes based on prior studies~\cite{miller2022did,sultana2022identification,cohen2021contextualizing,murphy2022pushback,rahman2024words,ferreira2022heated}. These attributes represent a participant's GitHub tenure, project experience, gender, and communication history. We compute each attribute for both the author and the target of a comment; therefore, we have 12 attributes from this category.

\subsection{Attribute Calculation}

To investigate RQ3 and RQ4, it is necessary to compute attributes at pull request (PR) and comment levels, respectively. Given that our dataset comprises 16 M PRs and 101.5 M comments, calculating the PR and comment-level attributes listed in Table~\ref{tab:rq-attributes} 
for the entire dataset would be exceedingly time-consuming and resource-intensive. Therefore, we reduced the sample size for RQ3 and RQ4 by randomly selecting 385 projects from the three project groups (i.e., `PRF-L', `PRF-M', and `PRF-H'). 
We choose this sample size to satisfy a 5\% error margin and 95\% confidence interval~\cite{cochran1977sampling}. This sample of 1,155 projects includes 6.3 M PRs, 30 M comments, and 416 K users. We exclude gaming projects from this analysis since they have a higher prevalence of toxicity, and many such projects do not consider profanities offensive. Therefore, contexts and participants of toxicity among gaming projects are not representative of non-gaming ones.
We wrote Python scripts and MySQL queries to compute the 32 attributes listed in Table~\ref{tab:rq-attributes} based on their definitions. While most attributes are straightforward to calculate, five require additional heuristics, as defined in the following.

    \textbf{Gender: }
    We adopted a similar protocol to~\citet{sultana2022code} to automatically predict users' genders. We have used genderComputer~\cite{vasilescu2014gender} and Wiki-Gendersort~\cite{berube2020wiki} tools to resolve the gender from a user's name, preferred pronoun, and location if available. We have also downloaded a user's GitHub avatar and applied an automated human face detection model~\cite{goyal2017face}. Further, we used a pre-trained photo to gender-resolution model~\cite{eidinger2014age} to predict the user's gender. Conflicts between the two approaches were resolved by manually investigating users' profiles.
 Finally, we successfully resolved 75.4\% of the total users (92\% with full names). We only include gender-resolved users for RQ4. 

 \textbf {Project Member:}
 Following the recommendation of ~\citet{gousios2012ghtorrent}, we consider a user a project member if that user has write access (i.e., merged at least one PR or created an intra-branch PR) to the repository.

 \textbf{GitHub Tenure and Project Tenure:}
We compute a user's GitHub tenure at an event as the months between their account creation and the event's timestamp. Similarly, we calculate a developer's project tenure during each project interaction (e.g., commit, pull request, or comment). 

 \textbf {Newcomer:}
Following the definitions of prior studies~\cite{steinmacher2013newcomers,subramanian2020analyzing}, we consider a user as a newcomer to an OSS project until they have got their first PR accepted to this project.

\subsection{Regression Modeling}

Regression analysis offers a robust statistical method for examining the influence of one or multiple independent variables on a dependent variable \cite{foley2018regression}. Two categories of regression models are used: (i) \textit{Predictive analysis}: involves creating a formula to forecast the value of a dependent variable based on the values of one or more independent variables; and (ii) \textit{Inferential analysis}: seeks to establish whether a specific independent variable affects the dependent variable and to quantify that impact if present \cite{allison2014prediction}. An inferential analysis differs from a predictive analysis in two key aspects. First, \emph{multicollinearity}: when two or more independent variables are highly correlated, incorporating all correlated variables simultaneously in inferential analysis can lead to an overfitting problem. However, in predictive analysis, multicollinearity is not a concern. Second, the importance of $R^2$ -- the goodness of fit of a regression model \cite{helland1987interpretation}. While a higher \emph{$R^2$} is desirable, it holds greater importance in predictive analysis. In inferential analysis, even with a low $R^2$, the regression model can provide valuable insights into the relationships between the independent and dependent variables \cite{allison2014prediction}. 
We train multivariate inferential regression models to analyze associations between the toxicity and the 32 attributes listed in Table~\ref{tab:rq-attributes}. The following subsections detail the regression models to answer RQ2, RQ3, and RQ4.

\subsubsection{Multinomial Logistic Regression for RQ2}
We found training a regression model for RQ2 challenging since computing various project characteristics variables at the creation timestamp of a comment requires the entire event log for a project (e.g., when a new star was added), which is resource-intensive to mine due to the enormous size of our dataset. While Google's BigQuery hosts a dataset of  GitHub events, it would be expensive to query this service. Therefore, we used aggregated attributes over the lifetime of a project. We calculated toxicity per hundred comments ($percent\_toxic$) for each project over its lifetime and used it as the dependent variable for RQ2. However,  if the dependent variable is a ratio, a model can identify spurious associations~\cite{richard-spurious}. 
Following the recommendation of ~\citet{long2006regression}, we transform the $percent\_toxic$ variable into a three-level categorical variable named $toxicity\_group$. 
We selected the number of categories and thresholds for this grouping based on the inflection points\footnote{points of a curve at which a change in the direction of curvature occurs} in the cumulative distribution curve.
The `Low toxic' group includes 324 projects with $percent\_toxic < $ 0.02\%. The 2,082 projects from the `Medium toxic' group have  0.02 $ \leq percent\_toxic < $ 1\%. The remaining 421 projects belong to the `High toxic' group with $percent\_toxic \geq $ 1\%. 
Since $ toxicity\_group$ has three levels, we use a Multinomial Logistic Regression (MLR) model, where $ toxicity\_group$ is the dependent variable and 11 project characteristics attributes are independents.

\subsubsection{Bootstrapped Logistic Regression for RQ3 and RQ4}
For RQ3, the dependent variable is \textit{HasToxicComment}, set to 1 if a PR has at least one toxic comment and 0 otherwise.
For RQ4, we use participant attributes computed at the comment level as independents. We use \textit{isToxic} as the dependent, $1$ if the comment is toxic, and 0 otherwise. 
We train two models for RQ4, one with the author's attributes as the independents and the other with the target's attributes.
Since the dependents are binary for RQ3 and RQ4, we use Logistic Regression models. 
As the dataset of RQ3 and RQ4 consist of a rare binary outcome variable (i.e., \textit{HasToxicComment}, \textit{isToxic}), we use a bootstrapped regression modeling technique~\cite{xu2020applications}. In this technique, we choose a desired ratio between the minority and the majority. We randomly downsample the majority until the desired ratio is reached. We fit a logistic regression model with each bootstrapped sample, measure its fit, and compute regression coefficients. This process is repeated 100 times, and we record the results of each iteration in a dataset. We report median and 95\% confidence interval for model fit and regression coefficients.  We also explored various ratios between the minority and the majority and found that the model's goodness of fit (i.e., $R^2$) reduces with the increment of the majority's share. We chose a ratio of 1:10 since increasing the majority's share beyond that produced unreliable models in a few cases, according to the Log-likelihood test (\textit{lrtest}).

\subsubsection{Correlation and Redundancy Analysis}
For an inferential regression model, multicollinearity poses a threat to validity.  We used the variable clustering approach suggested by \citet{sarle1990sas} to identify multicollinearity. With this approach, we create a hierarchical cluster representation of independents using Spearman's rank-order correlation test \cite{statistics2013spearman}. As recommended by ~\citet{hinkle1998applied}, we set the cutoff value at $|\rho| = 0.7$ for the correlation coefficient. Only the explanatory variable with the strongest correlation with the dependent was chosen from a cluster of variables with $|\rho| \geq 0.7$.

\begin{table*}
    \caption{For the bootstrapped logistic regression models, model fit measured using Vealll-Zimmermann Psuedo $R^2$.  A 95\% confidence interval is also reported for $R^2$ values. All models are significantly better than null models ($p<0.001$).}
    \label{tab:model_fit}
    \centering
   \vspace{-10pt}
    \input{Sections/Tables/model-fit}
    \vspace{-10pt}
\end{table*}

\subsubsection{Model analysis}
We also use the Log-likelihood test (\textit{lrtest})  to assess whether a model significantly differs (Chi-Square, $p<0.05$) from a null model and can be reliably used for inference.
We evaluate each model's goodness-of-fit using Veall-Zimmermann Psuedo-$R^2$~\cite{veall1994evaluating} since prior research~\cite{smith2013comparison} found this measure having closer correspondence to ordinary least square $R^2$.  A higher $R^2$ value indicates a better fit.  
We use the Odds ratio (OR) to quantify the association between the dependent and independents and estimate effect size. 
For a binary independent (e.g., $isGame$), OR indicates the odds of an outcome if the independent variable changes from 0 to 1, while all other factors remain constant. For a continuous variable (e.g., project age), OR indicates an increase or decrease in odds for the dependent with one unit change in the factor. In simple terms, OR $>$1 indicates a positive association and vice versa. We use the p-value of the regression coefficient to assess the significance of an association, with $p<0.05$ indicating a statistical significance.
Table~\ref{tab:model_fit} shows goodness-of-fit measured with Pseudo-$R^2$  for the regression models trained for RQ3 and RQ4. Since we bootstrapped each model 100 times, we report median and 95\% confidence intervals for each model. The results of \textit{lrtest} indicate that all models are significantly better than a null model ($p<0.001$) and are reliable to infer insights to answer our RQs. However, although models for RQ4 are significant, they have low $R^2$ values, which we further explain in Section \ref{sec:res-rq4}.

%% file: Sections/Tables/combined_attributes.tex
\resizebox{\linewidth}{!}{    
     \begin{tabular}{|p{1.9cm}|p{5.5cm}|p{11.2cm}|} \hline
    \textbf{Variable}     & \textbf{Definition} & \textbf{Rationale}  \\ \hline
    \rowcolor[gray]{.8}
\multicolumn{3}{l}{ \textbf{RQ2: Project Characteristics}}\\ \hline

PR /month* & Average number of PRs per month.  &  Indicates the volume of development activity. Active projects may have a higher probability of toxic interactions~\cite{miller2022did}.  \\  \hline

issues /month & Average number of issues per month. & A higher number of bugs indicates the lack of quality, which may cause frustration among users and developers~\cite{miller2022did,raman2020stress}.    \\  \hline

commits /month & Average number of commits per month. & Commit is another indication of the volume of development activity. High activity may cause burnouts~\cite{raman2020stress}, and lack may cause frustration among users~\cite{miller2022did}.    \\  \hline

release /month & Average number of releases per month. & Frequent releases may satisfy the customers to decrease toxicity, and vice versa~\cite{costa2018impact}.  \\  \hline

issue resolution rate & Percentage of issues resolved. & Users may become frustrated due to issues affecting them not being resolved~\cite{miller2022did}.   \\  \hline

isCorporate*  & Whether the project is sponsored by a corporation. & Corporate projects may have less toxicity than non-corporate ones due to the consequences of HR policy violations~\cite{raman2020stress}.   \\  \hline

project age* & 
Total months since a project's creation. & Older projects showed more toxicity~\cite{raman2020stress}.    \\  \hline

member count & Number of users with write access. & Toxicity increases with community size due to diverse views and higher potential conflicts~\cite{basirati2020understanding}.    \\  \hline

isGame* & Whether the project is gaming or not. & Prior studies have found prevalence of toxicity among gaming communities~\cite{miller2022did,belskie2023measuring,paul2018toxic,beres2021don}.    \\  \hline

stars & Number of stars on GitHub project. &  Popularity shows users' interests. Scrutiny and expectations increase with popularity and therefore stress on developers~\cite{raman2020stress}.   \\  \hline

forks & Number of forks on GitHub project. & Fork is another measure of project popularity~\cite{zhou2020has}. \\  \hline

\rowcolor[gray]{.8}
\multicolumn{3}{l}{ \textbf{RQ3: PR Context}}\\ \hline

commit count & Number of commits in a PR. & A large number of commits increases review effort~\cite{thongtanunam2017review}, which may frustrate reviewers, cause delays, and frustrate the author.  \\  \hline

number of changed files & The number of files changes in each PR. & A higher number of file changes requires a longer review time~\cite{review_2} and comprehension difficulty.
  
\\  \hline

code churn (log)* & The total number of rewritten or deleted code. & A high number of changed lines increases the probability of defects in the code~\cite{nagappan2005use,nagappan2007using} and may link to more toxic comments.  \\  \hline

isAccepted* & Whether the code review is accepted or rejected. & Developers used more toxic comments in rejected codes/patches~\cite{ferreira2021shut}.     \\  \hline

isBugFix & Whether the code review is for fixing a bug or not. & Issue discussions may instigate toxicity when the resolution is not liked by affected parties~\cite{ferreira2022heated,miller2022did}.    \\  \hline

change entropy (log) & A measure of change complexity, which estimates how much dispersed a changeset is among multiple files~\cite{thongtanunam2017review}. & Complexity of code change affects review time and participation~\cite{thongtanunam2017review}. Moreover, unnecessary complexity may be a sign of a poor quality change, which may receive harsh critique~\cite{rahman2024words}. \\  \hline

review interval & Time difference from the start of the code review to the end. & Delayed code reviews are more likely to cause frustration for developers~\cite{egelman2020predicting,turzo2023makes}.     \\  \hline

number of iterations (num iter)  & Total number of iterations (i.e., number of times changes requested) in a PR. &  Higher number of iterations frustrates both developers and reviewers due to additional time~\cite{turzo2023makes}. Higher iteration also indicates the lack of common understandings~\cite{ebert2019confusion} and potential disagreements~\cite{murphy2022pushback}.   \\  \hline

review comments* & The total number of review comments from reviewers in a PR.  & 
A higher number of review comments indicate significant concerns from the reviewers over its quality, which often causes toxicity~\cite{rahman2024words}.\\  \hline

\rowcolor[gray]{.8}
\multicolumn{3}{l}{ \textbf{RQ4: Participants}}\\ \hline

isWoman& Whether the person is a woman & Prior studies have found women and marginalized minorities as frequent victims of toxicity~\cite{raman2020stress, gunawardena2022destructive}.
    \\  \hline

isMember* & Whether the person is a project member or not. & Project members are the authors of many toxic comments in replying to the outside members' query~\cite{cohen2021contextualizing, miller2022did}.  \\  \hline

 isNewComer* & Whether the person is a newcomer to the current project. & Newcomers may get frustrated due to delays~\cite{steinmacher2013newcomers} and unfavorable decisions~\cite{ferreira2021shut}. \\  \hline 

 GitHub tenure* & Age of GitHub account, in terms of the {total} months at the time of an event. & \cite{miller2022did} reported toxic comments from accounts with no prior activity on GitHub. \\  \hline  

  project tenure & Tenure with the current project in terms of the {total} months. & Although long-term members of a project are more committed to maintaining a professional environment in a community, Miller \textit{et} al. found toxic comments from them~\cite{miller2022did}. Moreover,  they may be targets if their decisions are not liked by issue reporters~\cite{ferreira2022heated}.   \\  \hline 

    toxicity / month* & The total number of toxic comments a user posts per month. & ~\cite{miller2022did} found many repeat offenders, as many OSS developers have toxic communication styles~\cite{miller2022did,leaving-for-toxicity}.   \\  \hline

    \end{tabular}
    }

%% file: Sections/Tables/model-fit.tex

    \begin{tabular}{|l|c|c|c|} \hline

  {\textbf{Model}} &  \multicolumn{1}{c|} {\textbf{PRF-L}} &  \multicolumn{1}{c|}{\textbf{PRF-M}}& \multicolumn{1}{c|}{\textbf{PRF-H}} \\  \hhline{----}
  
   {RQ3} &  0.16 [0.15, 0.17] &    
   0.19 [0.19, 0.20] &     0.18 [0.18, 0.19]   \\  \hline

   {RQ4 (author)} &  0.01 [0.01 ,0.01] &    
   0.02 [ 0.02, 0.02] &      0.09 [0.09, 0.09]   \\  \hline

   {RQ4 (target)} &  0.01 [0.01 ,0.01] &   
   0.01 [ 0.01, 0.01] &     0.11 [0.11, 0.11]  \\  \hline

    \end{tabular}

%% file: Sections/results.tex
\section{Results}
\label{sec:results}

The following subsections detail the results of the four RQs.

\subsection{RQ1: Nature of toxicity}
\label{sec:results_rq1}
\begin{table*}
    \caption{The most common forms of toxicities with definitions within our sample of manually labeled 532 PR comments. We also showed the mapping from existing works.}
\label{tab:toxicity_types}
    \centering
\input{Sections/Tables/toxicity_types.tex}
\end{table*}

Table~\ref{tab:toxicity_types} shows the distributions of 11 forms of toxicities among our manually labeled dataset of 532 PR review comments.
Similar to ~\citet{miller2022did}'s investigation, we found profanity (i.e., severe language, swearing, cursing) as the most common form, with more than half of the sample ($\approx 58\%$) belonging to it. 
Toxic texts authored by OSS contributors frequently include profane words such as `shit', `fuck', `ass', `crap', `suck', and `damn'.
We found trolling to be the second most common form, with 18\%, followed by insult, self-deprecation, and object-directed toxicity. 
Identity attacks, insults, and threats, which are regarded as severe toxicities~\cite{goyal2022your}, were found among $\approx$22\% of the samples.
We also noticed over two-thirds of our samples $\approx$72\% belonging to multiple forms.  For example, \textcolor{brown}{“:laughing: Holy shit, you are fucking stupid. It is an extremely simple proc with a switch. Seriously, did you even look at the code? Next, you'll tell me all switches are copypaste.}” represents both profanity and insult. 
While toxicity on social media has higher occurrences of flirtation and obscenity~\cite{gunasekara2018review,goyal2022your}, we found $\approx$2\% such cases in our sample.


\begin{boxedtext}
\textbf{Key finding 1:} \emph{Profanity is the dominant form of toxicity in GitHub PRs. Severe toxicities, such as insults, identity attacks, and threats, represent $\approx$ 22\% cases. Unlike other online mediums, flirtation and obscenity were less common in our sample. }
\end{boxedtext}


\subsection{RQ2: Project characteristics}
\label{sec:rq2}
\begin{table}

    \caption{Results of our MLR model to identify associations of project characteristics with toxicity. We set the `Low toxic' group as the reference to compute odds ratios. Hence, $OR>1$ indicates a higher likelihood of a project transitioning to the `Medium' or `High'  toxic group with an increment of that factor and vice versa. }
    \label{tab:toxicity_projects}
    \centering
\input{Sections/Tables/project_characteristics_new}

\end{table}

During RQ2's model training
two factors (i.e., forks and pulls per month) were dropped due to multicollinearity and were not included in our MLR. We estimated the fit of our MLR with NagelKerke $R^2$ =0.224. Our Log likelihood test results suggest that this model significantly differs ($\chi^2$ =545.16, $p<0.001$) from a null model. 
In addition to modeling the probability of a specific result based on a group of independent variables, MLR also enables the assessment of the probability of transitioning to a different dependent category from the current one when a specific independent variable changes~\cite{bayaga2010multinomial}. Hence, we set the `Low toxic' projects as the reference group in MLR and compute the odds of a project moving to the `Medium toxic' or `High toxic' group if one of the independents changes by a unit.  
Table~\ref{tab:toxicity_projects} shows the result of our MLR model, with OR values for each factor. 
Our results suggest that projects with corporate sponsorship ($isCorporate$) are significantly less likely to belong to the `Medium toxic' or `High toxic' groups than the `Low toxic' group. HR rules, professional codes of conduct, and the potential for job loss may be the reasons. 
We also noticed a significantly higher level of toxicity among the popular projects ( i.e., stars). 
We found that the prevalence of toxicity significantly increased with project age. 
{Moreover, our analysis} found no significant association between toxicity and development activities (i.e., commits/month and releases/month) and project quality (i.e., issues/month). On the other hand, issue resolution rates (i.e., percentage of resolved issues) significantly reduce toxicity, as projects with higher rates are more likely to belong to the `Low toxic' group than the `Medium toxic' or `High toxic' group.
Supporting observations from prior studies~\cite{miller2022did}, we also noticed the significantly higher prevalence of toxicity among gaming projects, as a gaming project is seven times more likely to belong to the `High toxic' group than the `Low toxic' or `Medium toxic' group.

\begin{boxedtext}
\textbf{Key finding  2:} \emph{ While popularity and staleness are positively associated with the prevalence of toxicity, issue resolution rate has the opposite association.
While corporate-sponsored projects are likelier to be 'low toxic,' gaming projects are likelier to belong to the opposite spectrum. 
}
\end{boxedtext}


\subsection{RQ3: Pull request context}
\label{sec:rq3}
\begin{table}
    \caption{{Associations between pull request contexts and toxicity. Values represent the median odds ratio for each factor with 95\% confidence intervals inside brackets.}}
    \label{tab:toxicity_context}
    \centering
   
    \input{Sections/Tables/context_toxicity_updated}
\end{table}

One of the nine pull request context factors (i.e., the number of changed files) was dropped due to multicollinearity. Table~\ref{tab:toxicity_context} shows median odds ratios with 95\% confidence intervals for the remaining eight factors based on our bootstrapped logistic regression models repeated over 100 times. All eight factors show significant associations for the PRF-H group (i.e., PR/month $ >32$). For this group, bug fix PRs, code churn, review interval, the number of review comments, the number of review iterations, and change entropy are positively associated with toxicity. On the other hand, the number of commits and acceptance decisions are negatively associated. Similarly, we noticed almost identical associations for the PRF-M group (i.e., 8 $<$ PR/month $ <32$), except $isBugFix$ does not have a statistically significant association. For the PRF-L group (i.e.,  PR/month $ <8$), only three factors have statistically significant associations with toxicity, where the review interval and the number of review comments have positive ones. In contrast, $isAccepted$ has a negative one.

A positive correlation between toxicity and isBugFix for the PRF-H group indicates that discussions on approach to fix pending issues might become heated for highly active projects, which support prior studies on locked issues~\cite{ferreira2022heated,miller2022did}. However, such a trend is not seen among projects belonging to PRF-M and PRF-L. While the number of commits included in a PR is negatively associated with toxicity for both PRF-M and PRF-H groups, our results suggest contradicting associations between toxicity and commit size measured using code churn. Since we code churn and the number of commits included in a PR are positively correlated, we were surprised by this finding. {Our in-depth investigation suggests that PRs with large numbers of commits are often due to inter-branch clean-up or feature imports.}
Hence, such PRs are less likely to have discussions~\cite{thongtanunam2017review} and, therefore, are less likely to be toxic. Positive correlations between toxicity and review interval across all project groups suggest that delayed decisions frustrate the participants and, hence, are more likely to instigate toxicity. 
Similarly, the number of review comments for a PR, an indicator of issues identified by reviewers, is positively associated with toxicity. This result indicates that PRs with poor-quality code are more likely to be associated with toxicity.
Unsurprisingly, we found a negative correlation between accepted PRs and toxicity among all groups, which indicates that rejected PRs are significantly more likely to be associated with toxicity than accepted ones.
The number of iterations, which indicates the number of times an author must add additional commits based on reviewers' suggestions, is positively associated with toxicity for projects belonging to the PRF-M and PRF-H groups. 
Finally, change entropy, a proxy for change complexity, is positively associated with toxicity among projects belonging to PRF-M and PRF-H groups.

\begin{boxedtext}
\textbf{Key finding  3:} \emph{Accepted PRs are less likely to encounter toxicity. On the contrary,  code churn, review intervals, the number of review comments,  change entropy,  and the number of review iterations are positively associated with toxicity on GitHub. }
\end{boxedtext}


\subsection{RQ4: Participants}
\label{sec:res-rq4}
\begin{table*}
    \caption{{Associations between characteristics of authors toxicity. Values represent the median odds ratio for each factor with 95\% confidence intervals inside brackets.}}
    \label{tab:toxicity_authors}
    \centering
    \input{Sections/Tables/author_toxicity}
\end{table*}


\begin{table*}
    \caption{{Associations between characteristics of targets and toxicity. Values represent the median odds ratio for each factor with 95\% confidence intervals inside brackets.}}
    \label{tab:toxicity_targets}
    \centering
\input{Sections/Tables/target_toxicity}
\end{table*}

We train two types of regression models, one to investigate the characteristics of persons authoring toxic comments and the other with the targets. Similar to the RQ3, we train bootstrapped logistic regression models repeated over 100 times. Tables~\ref{tab:toxicity_authors} and ~\ref{tab:toxicity_targets} show the odds ratios of authors and targets for each factor with 95\% confidence intervals for the three PRF-based project groups. The results of Log-likelihood ratio tests ($lrtest$) suggest that these models are significantly better than Null models and, therefore, are suitable to provide inferential insights. However, these models have low $R^2$ (i.e., low explainability power). This result indicates that the characteristics of participants, i.e., the attributes used to fit these regression models, have a very low explanatory power for toxic occurrences. Regardless, our models indicate several participant characteristics having significant associations, which we detail in the following.

\begin{boxedtext}
\textbf{Key finding  4:} \emph{
Although occurrences of toxic comments are significantly associated with several participant characteristics, these have low explanatory power for toxicity in OSS PR contexts.}
\end{boxedtext}

\vspace{10pt}

\noindent \textbf{Characteristics of authors of toxic comments:}
Our results suggest significantly lower odds of women authoring toxic comments among PRF-L and PRF-H groups. Similarly, newcomers have lower authoring odds among PRF-M and PRF-H groups.
Among all three groups, project members are significantly less likely to author toxic comments, and the likelihood of being such an author significantly decreases with GitHub tenure.
The likelihood of authoring toxic comments significantly increased with project tenure among PRF-L and PRF-H groups. 
Our results support ~\citet{miller2022did}'s observation that there are many repeat offenders since the likelihood of authoring toxic comments significantly increases with the prior frequency of such occurrences.

\vspace{4pt}
\noindent \textbf{Characteristics of targets of toxic comments:}
Contrary to our expectations, formed based on results~\cite{gunawardena2022destructive,raman2020stress,steinmacher2015social}, we did not find any significantly higher odds of women or newcomers being targets of toxicity on GitHub. We noticed the opposite among PRF-H and PRF-L. Similarly, newcomers have significantly lower odds of becoming targets among PRF-H and PRF-M.
Being a project member significantly increases the odds of being a target among PRF-L and PRF-M but reduces among PRF-H.
Project tenure increases the odds of being a target for PRF-M and PRF-H but reduces among PRF-L.
The age of a GitHub account is positively associated with being a target only for PRF-M. 
Finally, these results suggest a `quid pro quo,' i.e., prior frequent authoring of toxic comments significantly increases the odds of becoming a target.


\begin{boxedtext}
\textbf{Key finding  5:} \emph{Women and newcomers are less likely to be either authors or targets of toxic comments in GitHub PR comments. Developers who have authored toxic comments frequently in the past are significantly more likely to repeat and more likely to become toxicity targets. }
\end{boxedtext}


%% file: Sections/Tables/toxicity_types.tex
\resizebox{\textwidth}{!}{  
    \begin{tabular}{|p{1.8cm}|p{3.6cm}|p{3.8cm}|p{4cm}|>{\raggedleft\arraybackslash}p{1cm}|r|} \hline

     \textbf{Type} & \textbf{Mapping}& \textbf{Definition}     & \textbf{Example} &  \textbf{Count$\ddag$}& \textbf{Ratio}\\ \hline
     
Profanity & Profanity~\cite{sarker2022automated}, Expletives~\cite{miller2022did}, Vulgarity~\cite{ferreira2021shut}  & A comment that includes profanity. & \textcolor{brown}{``You know, at some point, github fucked me over. In Visual Studio, this was just fine, wtf....''}  &  311 & $58.45\%$  \\  \hline

Trolling & Trolling~\cite{miller2022did, ferreira2021shut} & Using trolling with destructive discussions and those are more severe and provoke arguments. &  \textcolor{brown}{``$@$clusterfuck There's a difference between being in cryogenics and not in cryogenics you big nerd''} &96 & 18.04\%\\  \hline

Insult & Insult~\cite{miller2022did,sarker2022automated} & Disrespectful expression towards another person. &  \textcolor{brown}{``Acknowledge that the vote wasn't entirely singulo shitposters $>$ ARE YOU SCHIZOPHRENIC??  Jesus dude why are you even here still''}  &92 & $17.3\%$ \\  \hline

Self-deprecation & Self-deprecation~\cite{miller2022did} & If a demeaning word towards him/herself consists of severe language, it would be marked as self-deprecation.  &  \textcolor{brown}{``@ComicIronic Okay, I'll fix it when I fix my shitty code, which will  have to happen tomorrow''} &67 & 12.6\% \\  \hline

Object Directed Toxicity & {New} & Anger, frustration, or profanity directed
toward software, products, or artifacts.  & \textcolor{brown}{``the PR has fallen into conflict hell, I'll be closing this and re-opening some of its changes shornestly''} &  49 &9.21\% \\ \hline

Entitled & Entitled~\cite{miller2022did} & When people demand due to the expectation related to contractual relationship or payment.  & \textcolor{brown}{``Again I didn't break it  Are you fucking stupid lol  $>$ merge your update into PR $>$ buckling doesn't work''} & 17& 3.2\% \\  \hline

Identity attack & Identity attack~\cite{sarker2022automated}, Inappropriate jokes about an employee~\cite{egelman2020predicting} &  Attacking the person's identity. & \textcolor{brown}{``Fuck those argentinians. Did you test it?''} &17 & 3.2\% \\  \hline

Threats & Threats~\cite{ferreira2021shut,sarker2022automated} & A behavior that is aggressive or threatening someone. & \textcolor{brown}{``Done - I can always revoke your access if you mess things up ;''} &12 & 2.25\% \\  \hline

Obscenity & Reference to sexual activities~\cite{sarker2022automated} &An extremely offensive comment that demeans women or LGBTQ+ people.  & \textcolor{brown}{``dude you need to spend less time on programming and more time with women''} &6& $1.1\%$ \\ \hline 

Arrogance &Arrogance~\cite{miller2022did} & Imposing the own view on others due to superiority. & \textcolor{brown}{``Araneus is shit and generic as hell I think steely is an acceptable name''} &5 & $<1\%$ \\  \hline

Flirtation &Flirtation~\cite{sarker2022automated} & A comment that represents flirting. &  \textcolor{brown}{``Frigging love you Niki. Seriously''} &5 & $<1\%$ \\ \hline 

\multicolumn{6}{l}{$\ddag$ -\textit{since a text can belong to multiple categories, the sum of the categories is greater than our sample size.}}

    \end{tabular}
}

%% file: Sections/Tables/project_characteristics_new.tex

\begin{tabular}{l|r|l|r|l}
\hline
\multicolumn{1}{c|}{\multirow{2}{*}{\textbf{Attribute}}} & \multicolumn{2}{c|}{\textbf{Medium toxic}} & \multicolumn{2}{c}{\textbf{High toxic}} \\ \cline{2-5} 
\multicolumn{1}{c|}{}   & \multicolumn{1}{c|}{\textbf{OR}}    & \multicolumn{1}{c|}{\textbf{$p$}}   & \multicolumn{1}{c|}{\textbf{OR}}                                 & \multicolumn{1}{c}{\textbf{$p$}}   \\ \hline

isCorporate  & 0.888 & \textbf{0.000}$^{***}$ & 0.47 & \textbf{0.000}$^{***}$ \\  \hline
member count   & 0.999 & 0.821 & 1.0001 & 0.754  \\  \hline
stars & 1.001 & \textbf{0.000}$^{***}$ & 1.001 & \textbf{0.000}$^{***}$ \\  \hline
issues/month & 1.001 & 0.234 & 0.998 & 0.061 \\  \hline
project age &  1.004 & \textbf{0.000}$^{***}$ & 1.009 & \textbf{0.000}$^{***}$ \\  \hline
commits/month & 1.0001 & 0.316 & 1.0001 & 0.447 \\  \hline
release/month   & 1.003 & 0.310 & 0.998 & 0.813  \\  \hline
bug resolution   & 0.204 & \textbf{0.000}$^{***}$ & 0.985 & \textbf{0.000}$^{***}$ \\  \hline

isGame   &  0.486 & \textbf{0.000 }$^{***}$ &  7.259 & \textbf{0.000}$^{***}$ \\  \hline

\multicolumn{5}{l}{*** , **, and *  represent statistical significance at $p <$ 0.001, }\\

\multicolumn{5}{l}{ $p <$ 0.01, and $p <$ 0.05 respectively.}

\end{tabular}

%% file: Sections/Tables/context_toxicity_updated.tex
\resizebox{.9\textwidth}{!}{    

    \begin{tabular}{|p{3 cm}|R{3 cm}|R{3 cm}|R{3 cm}|} \hline
    
     \textbf{Attribute}     & \multicolumn{1}{c|}{\textbf{PRF-L}} & \multicolumn{1}{c|}{\textbf{PRF-M}} & \multicolumn{1}{c|}{\textbf{PRF-H}} \\ 
     \hline
     


isBugFix & 1.01[0.99, 1.05] & 1.02 [1.01, 1.03] & 1.06*** [1.06, 1.07]  \\  \hline

commit count & 1.00 [1.00, 1.01] & 0.99* [0.99, 0.99] & 0.99*** [0.99, 1] \\  \hline

code churn (log) &  1.11 [1.10, 1.12] & 1.13*** [1.13, 1.14] &  1.11*** [1.11, 1.11]\\  \hline

 review interval& 1.11*** [1.11, 1.12] & 1.17*** [1.17, 1.17] & 1.20***[ 1.20, 1.20]  \\  \hline
 
review comments  & 1.06*** [1.05, 1.07] & 1.06*** [1.05, 1.06] & 1.04*** [1.04, 1.04] \\  \hline

isAccepted & 0.77*** [0.75, 0.80] & 0.71*** [ 0.70, 0.73] & 0.93*** [0.93, 0.94] \\  \hline

num iter & 1.01 [0.99, 1.03] & 1.01*** [1.01, 1.02] & 1.01*** [1.01, 1.01] \\  \hline

change entropy (log) & 1.57 [1.51, 1.67] & 1.35*** [1.32, 1.37] & 1.26*** [1.25, 1.27]  \\  \hline

\multicolumn{4}{p{14cm}}{ *** , **, and *  represent statistical significance at $p <$ 0.001, $p <$ 0.01, and $p <$ 0.05 respectively.}


    \end{tabular}
}

%% file: Sections/Tables/author_toxicity.tex
\resizebox{.9\textwidth}{!}{    


   \begin{tabular}{|p{3 cm}|R{3 cm}|R{3 cm}|R{3 cm}|} \hline
    
     \textbf{Attribute}     & \multicolumn{1}{c|}{\textbf{PRF-L}} & \multicolumn{1}{c|}{\textbf{PRF-M}} & \multicolumn{1}{c|}{\textbf{PRF-H}} \\ 
     \hline

isWoman & 0.80** [ 0.71, 0.793] & 1 [0.96, 1.01] & 0.90*** [0.86, 0.87] \\  \hline

isNewComer & 1.09 [ 1.05, 1.14] & 0.88*** [0.86, 0.89] & 0.69*** [0.68, 0.69] 
 \\  \hline

isMember  & 0.89** [0.87, 0.92]  & 0.77*** [0.77, 0.88] & 0.64***  [0.64, 0.64] \\  \hline

 GitHub tenure&   0.99*** [ 0.99, 0.99] & 0.99*** [0.99, 0.99] & 0.99*** [0.99, 0.99]  \\  \hline
 
project tenure  & 1.01** [1.01, 1.01]   & 1.01 [0.99, 1.01] & 1.01***  [1.01, 1.01] \\  \hline

toxicity/month  & 1.06*** [1.05, 1.07] & 1.02*** [1.02, 1.02] & 1.01*** [1.01, 1.01]  
 \\  \hline

\multicolumn{4}{l}{ *** , **, and *  represent statistical significance at $p <$ 0.001, $p <$ 0.01, and $p <$ 0.05 respectively.}


    \end{tabular}
}

%% file: Sections/Tables/target_toxicity.tex
\resizebox{.9\textwidth}{!}{    


   \begin{tabular}{|p{3 cm}|R{3 cm}|R{3 cm}|R{3 cm}|} \hline
    
     \textbf{Attribute}     & \multicolumn{1}{c|}{\textbf{PRF-L}} & \multicolumn{1}{c|}{\textbf{PRF-M}} & \multicolumn{1}{c|}{\textbf{PRF-H}} \\ 
     \hline


isWoman & 0.48*** [ 0.46, 0.50] & 0.95 [ 0.93, 0.98] & 0.86*** [ 0.85, 0.87] \\  \hline

isNewComer & 0.91 [0.88, 0.95] & 0.90***[0.89, 0.92] & 0.74*** [0.74, 0.75] \\  \hline

isMember  & 1.12* [1.09, 1.15]  & 1.04 [1.02, 1.05] & 0.84*** [0.84, 0.84] \\  \hline

 GitHub tenure & 1.01 [0.99, 1.01] & 1.01*** [1.01, 1.01] & 1.01 [1.01, 1.01]  \\  \hline
 
project tenure & 0.99* [0.99, 0.99] & 1.01***[ 1.01, 1.01] & 1.01*** [1.01, 1.01]\\  \hline

toxicity/month & 1.99*** [ 1.81, 2.19] & 1.46*** [1.36, 1.52] & 1.26*** [1.25, 1.26]  \\  \hline

\multicolumn{4}{l}{ *** , **, and *  represent statistical significance at $p <$ 0.001, $p <$ 0.01, and $p <$ 0.05 respectively.}


    \end{tabular}
}

%% file: Sections/discussion.tex
\section{Discussion}
\label{sec:discussion}
The following subsections compare our findings against prior works and suggest recommendations.

\subsection{Potential Explanations of Several  Key Findings}
The results of our RQ2 (Section~\ref{sec:rq2} suggest that project popularity, measured in terms of the `number of stars,' is associated with increased toxicity. A project's popularity may put pressure on contributors to deliver new features and maintain quality. However, a rapid development pace can cause stress, burnout, and toxicity. We also found toxicity increasing with project age. Our manual investigation of sample projects suggests staleness (i.e., lack of response to issues or PRs) may be a potential reason.

The results of RQ3 (Section~\ref{sec:rq3}) suggest that review duration and the number of required iterations are positively associated with toxicity, with stronger associations seen among higher PRF groups. {These results suggest that frustrations may grow between authors and reviewers due to multiple review iterations, particularly among projects with higher activity levels.}
We also found a positive association between code complexity and toxicity. {This result indicates that complex changes, which are difficult to understand and review, may cause confusion~\cite{ebert2019confusion} and are more likely to be associated with toxicity.}

\begin{table*}
    
    \caption{Comparison against prior empirical studies investigating anti-social behaviors among OSS project}
    \label{tab:comparison}
    \centering   
\input{Sections/Tables/comparison_with_other_studies}
\end{table*}

\subsection{Comparison with  Prior SE Studies} 

Our large-scale empirical investigation includes 32 attributes from four categories. Out of those, 11 were investigated in other contexts in prior studies. Therefore, Table~\ref{tab:comparison} compares sample size, projects, and the number of factors and overlaps with our studies against the others to illustrate the novelty and significance of this study. Similar to~\citet{miller2022did}, profanity is the prevalent toxicity in our randomly sampled dataset. They reported a high share (25\%) of entitled issue comments, which are demands to project maintainers as if they had a contractual relationship or obligation~\cite{miller2022did}. However, we found only 3.2\% such cases in our sample. Supporting their findings, our results indicate repeat offenders, toxicity increasing with project popularity, long-term project contributors being authors of toxicity, and gaming projects harboring more toxic cases~\cite{miller2022did}. They also reported toxic comments from new GitHub accounts~\cite{miller2022did}. Aligning with this finding, we noticed the likelihood of authoring toxic comments decreased with GitHub tenure.
However, contrary to their findings, we notice a lower likelihood of project newcomers authoring toxic comments.
Our results also concur with one finding by~\citet{raman2020stress}, as we found a lower likelihood of toxicity among corporate-sponsored projects.
During their manual investigation of the Linux kernel, \citet{ferreira2021shut} found uncivil comments during reviews of rejected codes. Aligning with their findings, we noticed lower odds of toxicity among accepted PRs. They also reported incivility among project maintainers' feedback~\cite{ferreira2021shut}. However, contrasting their findings, we noticed a lower likelihood of toxicity from project members.
\citet{egelman2020predicting} reported a higher likelihood of pushback on large code changes. Our result aligns with this finding, as we found that the odds of toxicity increase with code churn.
\citet{raman2020stress} reported incivility due to poor-quality code changes. While we did not measure code quality directly, we may use the number of review comments as an indication of code quality since each review comment indicates an issue identified by a reviewer. Aligning with their findings, our results indicate higher odds of toxicity with the number of review comments.

\vspace{3pt}
\noindent \textbf{Potential reasons behind some of our findings contradicting prior studies:}
While we do not have a concrete answer to why some of our results contradict prior studies, we hypothesize that sampling differences may be a major factor. Prior studies picked samples from contexts where antisocial behaviors are more likely to occur (e.g., locked issues, rejected patches, heated discussions). However, these cases are not very frequent. For example, only 3.1\% of PR-linked issues in our sample are locked.
We noticed the most differences (i.e., two) against~\citet{miller2022did}, an exploratory study investigating a sample of 100 locked issues. Although valid for a specific context, their {selected} cases may not represent broader trends across GitHub. For example, arrogant contributors forcefully demanding acceptance of their pull request is a reasonable cause to lock issue threads. Hence, they obtained 25\% entitlement, but such cases are significantly lower among non-locked issue threads. For the same reason, \citet{miller2022did} reported {entitled type} behavior from newcomers, but our analysis suggests that newcomers are less likelier to author toxic texts than long-term contributors.
We also noted a discrepancy compared to ~\citet{ferreira2021shut}, who drew their sample from rejected patches of the Linux kernel mailing list and reported uncivil behavior from maintainers. Several Linux kernel maintainers have been known for their blunt communication style for years~\cite{toxic-blog-linux1,toxic-blog-linux3}. Our findings suggest that what was reported from LKML may not be a broader trend across the OSS spectrum. These contradictions also highlight the need for a large-scale study with diverse samples to understand the landscape of toxicity better.

\subsection{Actionable Recommendations}
Due to our study design, we cannot claim causal relationships for the associations identified in this study. However, some of the following recommendations apply only if such relationships exist.

\vspace{4pt}
\noindent \textbf{ I. Project Maintainers:}  
Our results from RQ3 (i.e., table~\ref{tab:toxicity_context}) suggest that delays in fixing bugs or answering user queries may create unhappy users and toxic comments targeted toward maintainers.
As a project's popularity grows, maintainers should focus on improving bug resolution since our results also show that a higher bug resolution rate is negatively associated with toxicity. Even if an issue is delayed, maintainers should respond politely and suggest workarounds, if possible, to avoid toxic interactions.
From the RQ4 analysis, project tenure is positively associated with toxicity. Therefore, building a positive culture needs to start with project maintainers since they are likelier to be project members with the longest tenures~\cite{toxic-blog-linux3}. 
Supporting prior studies~\cite{miller2022did,belskie2023measuring,paul2018toxic,beres2021don}, we also found a proliferation of toxicity among gaming projects in RQ2. Therefore,  we recommend that maintainers of gaming projects adopt a Code of Conduct and its enforcement mechanism to build a diverse community.

\vspace{4pt}
\noindent\textbf{II. Developers:}   We recommend developers avoid creating pull request contexts that are positively associated with toxicity.  For example, our results from RQ3 indicate that delayed pull requests are associated with toxicity. Therefore, reviewers should provide on-time reviews to avoid frustrating authors.
Similarly, large code changes are not only bug-prone~\cite{bosu2015characteristics} and difficult to review~\cite{thongtanunam2017review} but also likely to encounter toxicity. Therefore, when possible, creating pull requests with smaller changes is recommended. According to the findings from RQ3, pull requests with a large number of issues indicate poor quality codes and are more likely to receive harsh critiques. Therefore, developers should not create pull requests with changes that do not yet meet the quality standards for a project. A higher number of review iterations also frustrates authors and may cause toxicity. Hence, if possible, reviewers should request all required changes within a single cycle to avoid back and forth.
Complex changes are hard to review and are more likely to receive toxicity. Hence, authors should annotate such changes and include helpful descriptions to avoid confusion~\cite{ebert2019confusion} as well as toxicity.
 Even when frustrated or angry, developers should not use toxic languages since developers who use such languages are more likely to become victims (i.e., findings from RQ4). 
Finally, while contrary to prior evidence, we find that women and newcomers are less likely to be targets of toxicity in RQ4, we still recommend long-term contributors avoid such language if such persons are present in a discussion since toxicity not only dissuades newcomers from becoming a part of the communities~\cite{steinmacher2015social} but also disproportionately hurts minorities~\cite{gunawardena2022destructive}.

\vspace{4pt}
\noindent\textbf{III. Prospective joiners: } If a newcomer wants to avoid negative experiences associated with toxic cultures, we recommend they start with a corporate-sponsored OSS project that matches their expertise and interests. We also recommend such contributors avoid gaming or stale projects.

\vspace{4pt}
\noindent \textbf{IV. Researchers:}
We found variations among terminologies used for almost identical concepts among SE studies investing in anti-social behaviors. We also noticed conflicting opinions about whether a particular category should be considered anti-social. Since existing schemes are primarily based on the decisions of the respective researchers, they may not reflect the broader OSS community. Therefore, existing identification tools based on these schemes may not align with OSS developers' needs and would fail to achieve broader adoption. Moreover, recent research suggests whether a text should be considered toxic depends on various demographic characteristics~\cite{goyal2022your}. Hence, understanding the opinions of the broader OSS community and how their demographics influence perspectives of toxicity is essential to developing a custom mitigation strategy. 

%% file: Sections/Tables/comparison_with_other_studies.tex
\resizebox{\textwidth}{!}{  
    \begin{tabular}{|p{1.8cm}|p{2cm}|p{2.5cm}|p{3.5cm}|p{1.5cm}|p{3.5cm}|} \hline

     \textbf{Study} & \textbf{Method}  & \textbf{Sample Size}   & \textbf{{Sampling Criteria}} &  \textbf{\# factors}& \textbf{Comparison with ours}\\ \hline
     
Raman \textit{et} al.~\cite{raman2020stress} & Quantitative & 872k issues from 30 popular projects. & i) Training dataset from  `too heated locked issues; ii) Poor performance of their classifier with 47\% F1-score. &  3 & One overlapping factor, which concurs with our finding.  \\  \hline

Ferreira \textit{et} al.~\cite{ferreira2021shut} &	Qualitative &	1,545 email threads from Linux. &	i) Only rejected patches, ii) Linux kernel maintainers are known to be harsh.	& 9	& Two overlapping factors, where finding for one contradicts, and the other one concurs. \\  \hline

Miller \textit{et} al.~\cite{miller2022did} &	Qualitative &	100 issue discussions.	& i) Small sample, ii) Only locked issue threads.	& 10	& Six overlapping factors, where findings for two contradict, and the remaining four concur.

\\  \hline

Egelman \textit{et} al.~\cite{egelman2020predicting} &	Opinion survey &	Surveyed 1,397 developers in Google.	& i) One organization, ii) lack of quantitative validation	& 5	& Three overlapping factors, where all concur with our finding. \\  \hline

Ours &	Mixed, mostly quantitative &	2,828 GitHub projects and over 100M comments. &	Limitations of ToxiCR~\cite{sarker2022automated} applies. &	32	&  - \\  \hline

    \end{tabular}
}

%% file: Sections/threats.tex
\section{Threats to Validity}
\label{sec:threats}
\noindent \textbf{Internal Validity}
Our selection of 2,828 GitHub projects based on our sampling method threatens internal validity. GitHub hosts over 284 million projects, and mining all of them is infeasible. We defined six filtering criteria to reduce this sample space to 89k without excluding projects with significant communication and collaboration. A lower threshold for the number of contributors or stars would increase the number of projects in this sample and may potentially change our results. 
We applied a stratified sampling strategy to categorize the projects according to PR activity to encounter this threat. Therefore, threats due to threshold selection are more likely to influence only the PRF(L) group since most of the projects with a lower number of contributors or stars would fall under this group. However, there is no evidence that changing these thresholds would significantly alter the results, even for the PRF(L). Our selection of the list of attributes represents another threat to internal validity. Prior studies have found various factors such as politics or ideology triggering toxicity~\cite{miller2022did}. However, we could not investigate those factors due to the unavailability of automated mechanisms to identify such scenarios at a large scale.  This study only investigates automatically measurable factors that may be associated with toxicity.


\vspace{2pt} \noindent \textbf{Construct Validity}
Our \textit{(first)} threat in this category is due to using ToxiCR~\cite{sarker2022automated} to identify toxic comments automatically. 
Our validation of ToxiCR found 88.88\% precision, which is within the sampling error margin reported by ToxiCR's authors. ToxiCR has false positives in approximately one out of 10 cases. Similarly, ToxiCR has a false negative rate of between 10-14\%. Hence, these false positives and negatives may have influenced our results if ToxiCR is biased for/against any particular attributes (e.g., review interval or woman) included in our study. However, we do not have any evidence of such biases.
\textit{(Second)}, our manual labeling scheme to identify the nature of toxicities to answer RQ1 is a threat. Although multiple SE studies have studied antisocial behaviors, no agreed-upon scheme exists. Moreover, researchers from NLP and SE domains have used different terminologies to characterize similarly subjective concepts.
To mitigate this threat, we have analyzed existing studies~\cite{sarker2022automated,miller2022did, ferreira2021shut,egelman2020predicting,ferreira2022heated} and aggregated their categories to build our scheme. We acknowledge the subjectivity bias, where another set of researchers disagree with our scheme and definitions. \textit{(Third)}, our manual labeling process may have subjectivity biases. We prepared a scheme with category definitions and examples to mitigate this threat. The labelers had a discussion session before starting to build an agreed-upon understanding. We also measured inter-rater reliability to assess your labeling process. 
\textit{(Finally)}, automated gender resolution is another threat. We followed a procedure as the ones in multiple recent empirical studies~\cite{sultana2022code,santamaria2018comparison,bosu2019diversity}. We used multiple gender resolution tools, considered users' location and profile photos, and searched LinkedIn to improve resolution accuracy. This resolution process may be subject to misclassification. We did not attempt to identify
non-binary genders since we are unaware of any automated resolution of those without users' inputs.

\vspace{2pt} \noindent \textbf{External Validity}
The nature of toxicities in an OSS project may depend on factors such as project domain, governance, the number of contributors, and project age. 
We used a stratified random sampling strategy to select 2,828 projects representing diverse demographics, including the top OSS projects on GitHub, such as Kubernetes, Odoo, PyTorch, Rust, Ansible, pandas, rails, Django, numpy, angular, flutter, CPython, and node.js. 
Yet, our sample and its results may not adequately represent the entire OSS spectrum.

\vspace{2pt} \noindent \textbf{Conclusion Validity}
We assess the reliability of our models using goodness-of-fit metrics and log-likelihood tests. Hence, we do not anticipate any threats from the results obtained from our models.  Although our models account for various confounding variables, these models identified associations between dependents and predictors, and no causal relationships can be implied.

%% file: Sections/conclusion.tex
\section{Conclusion}

\label{sec:conclusion}
We conducted a large-scale mixed-method empirical study of 2,828 GitHub-based OSS projects to understand 
the nature of toxicities on GitHub and how various measurable characteristics of a project, a pull request's context, and participants associate with their prevalence.
We found profanity to be the dominant form of toxicity on GitHub, followed by trolling and insults. 
While a project's popularity is positively associated with the prevalence of toxicity, its issue resolution rate has the opposite association.
Corporate-sponsored projects are less toxic, but gaming projects are seven times more likely than non-gaming ones to have a high volume of toxicities. OSS developers who have authored toxic comments in the past are significantly more likely to repeat them and become toxicity targets.
Based on the results of this study and our experience conducting it, we provide recommendations to OSS contributors and researchers.

%% file: Sections/acknowledgement.tex
\section*{Acknowledgment}
\label{sec:acknowledgment}
This research is partially supported by the US National Science Foundation under Grant No. 
1850475 and 2340389 and funding from The University of Nebraska at Omaha, College of Information Science and Technology. 
The findings of this research do not necessarily provide the views of the National Science Foundation.
Jaydeb Sarker was affiliated with the Wayne State University when this research was conducted.

%% file: Sections/data_avail.tex
\section*{Data Availability}
\label{sec:data_avail}
We have made our dataset and source code publicly available at: \hyperlink{https://doi.org/10.5281/zenodo.14802294}{10.5281/zenodo.14802294}.